\documentclass[aps,prb,preprint,superscriptaddress]{revtex4-1}

\usepackage[dvips]{graphicx}
\usepackage{amssymb,amsfonts,amsmath}
\usepackage{color}
\usepackage{soul}

\newcommand\comment[1]{}

\newcommand{\myhl}[1]{#1}

\begin{document}

\title{Entropic Effect on the Rate of Dislocation Nucleation}

\author{Seunghwa Ryu$^1$, Keonwook Kang$^2$ and Wei Cai$^3$}

%
\affiliation{$^1$Department of Physics, Stanford University, Stanford, California 94305 \\
$^{2,3}$Department of Mechanical Engineering, Stanford University, Stanford, California 94305}

\begin{abstract}
Dislocation nucleation is essential to our understanding of
plastic deformation, ductility and mechanical strength of
crystalline materials.
Molecular dynamics simulation has played an important role in
uncovering the fundamental mechanisms of dislocation nucleation,
but its limited time scale remains a significant challenge for
studying nucleation at experimentally relevant conditions.
Here we show that dislocation nucleation rates can be accurately
predicted over a wide range of conditions by determining the
activation free energy from umbrella sampling.
Our data reveal very large activation entropies, which contribute
a multiplicative factor of many orders of magnitude to the
nucleation rate.
The activation entropy at constant strain is caused by thermal
expansion, with negligible contribution from the vibrational
entropy.
The activation entropy at constant stress is significant larger
than that at constant strain, as a result of thermal softening.
The large activation entropies are caused by anharmonic effects,
showing the limitations of the harmonic approximation widely used
\myhl{for rate estimation} in solids.
Similar behaviors are expected to occur in other nucleation
processes in solids.
\end{abstract}


\maketitle


Nucleation plays an important role in a wide range of physical,
chemical and biological
processes~\cite{ref:DWu,ref:Anderson,ref:Matsu,ref:Laaksonen,ref:Hanggi,
ref:tenWolde}.
%
In the last two decades, the nucleation of dislocations in
crystalline solids has attracted significant attention, not only
for the reliability of microelectronic devices~\cite{ref:Izumi},
but also as a responsible mechanism for incipient plasticity in
nano-materials~\cite{ref:LiNature,ref:LiMRS,ref:ZhuMRS} and
nano-indentation~\cite{ref:Schuh,ref:JLi,ref:Schall}.
%
%
However, predicting the nucleation rate as a function of
temperature and stress from fundamental physics is extremely
difficult.
Because the critical nucleus can be as small as a few lattice
spacings, the applicability of continuum theory~\cite{ref:Xu}
becomes questionable.
At the same time, the time scale of molecular dynamics (MD)
simulations is about ten orders of magnitude smaller than the
experimental time scale.
Hence MD simulations of dislocation nucleation are limited to
conditions at which the nucleation rate is extremely
high~\cite{ref:Bringa,ref:Tschopp}.

One way to predict dislocation nucleation rate under common
experimental loading rates~\cite{ref:ZhuPRL} is to combine the
transition state theory
(TST)~\cite{ref:Hanggi,ref:Vineyard} and the
nudged-elastic-band (NEB) method~\cite{ref:NEB}.
TST predicts that the nucleation rate per nucleation site in a
crystal subjected to constant strain $\gamma$ can be written as
\begin{equation}
 I^{\rm TST} = \nu_0 \, \exp\left[
             -\frac{F_c(T,\gamma)}{k_BT}\right]
 \label{eq:TST}
\end{equation}
where $F_c$ is the activation free energy, $T$ is temperature, and
$k_B$ is Boltzmann's constant.
The frequency prefactor is $\nu_0 = k_B T/h$, where $h$ is
Planck's constant.
Note that $F_c(T,\gamma) = E_c(\gamma) - T S_c(\gamma)$, where
$E_c$ and $S_c$ are the activation energy and activation entropy,
respectively.
Here we assume the dependence of $E_c$ and
$S_c$ on $T$ is weak, which is later confirmed numerically for $T \leq 400$K.
For a crystal subjected to constant stress $\sigma$,
$F_c(T,\gamma)$ in Eq.~(\ref{eq:TST}) should be replaced by the
activation Gibbs free energy $G_c(T,\sigma) = H_c(\sigma)- T
S_c(\sigma)$, where $H_c$ is the activation enthalpy.
Because the NEB method only computes the activation energy, the
contribution of $S_c$ is often ignored in rate estimates in
solids.
Recently, an approximation of $S_c(\sigma) = H_c(\sigma) / T_m$ is
used~\cite{ref:ZhuPRL}, where $T_m$ is the surface disordering
temperature. This approximation was questioned by subsequent MD
simulations~\cite{ref:Warner}.
%
%
\myhl{The magnitude of $S_c$ remains unknown because none of the
existing methods for computing activation free
energies}~\cite{ref:FS,ref:FTstring,ref:FFSFE} \myhl{has been
successfully applied to dislocation nucleation}.

We successfully applied the umbrella sampling~\cite{ref:FS} method
to compute the activation free energy for homogeneous and
heterogeneous dislocation nucleation in copper.
Based on this input, the nucleation rate is predicted using the
Becker-D\"oring theory~\cite{BD}.
%
Comparison with direct MD simulations at high stress confirms the
accuracy of this approach.
Both $F_c(T,\gamma)$ and $G_c(T,\sigma)$ show significant
reduction with increasing $T$, corresponding to large activation
entropies.
For example, $S_c(\gamma\!\!=\!\!0.092)\!=\!9\,k_B$ and
$S_c(\sigma\!\!=\!\!2\,{\rm GPa})\!=\!48\,k_B$ are observed in
homogeneous nucleation.
We found that $S_c(\gamma)$ is caused by the anharmonic effect of
thermal expansion, with negligible contribution from the
vibrational entropy.
The large difference in the two activation entropies, $\Delta S_c
\equiv S_c(\sigma) - S_c(\gamma)$, is caused by thermal softening,
which is another anharmonic effect.
Similar behaviors
%
%
are expected to occur in other nucleation processes in solids.
%

%
%
%
%
%
%

\begin{figure}[htb]
\begin{center}
\includegraphics[width=1.7in]{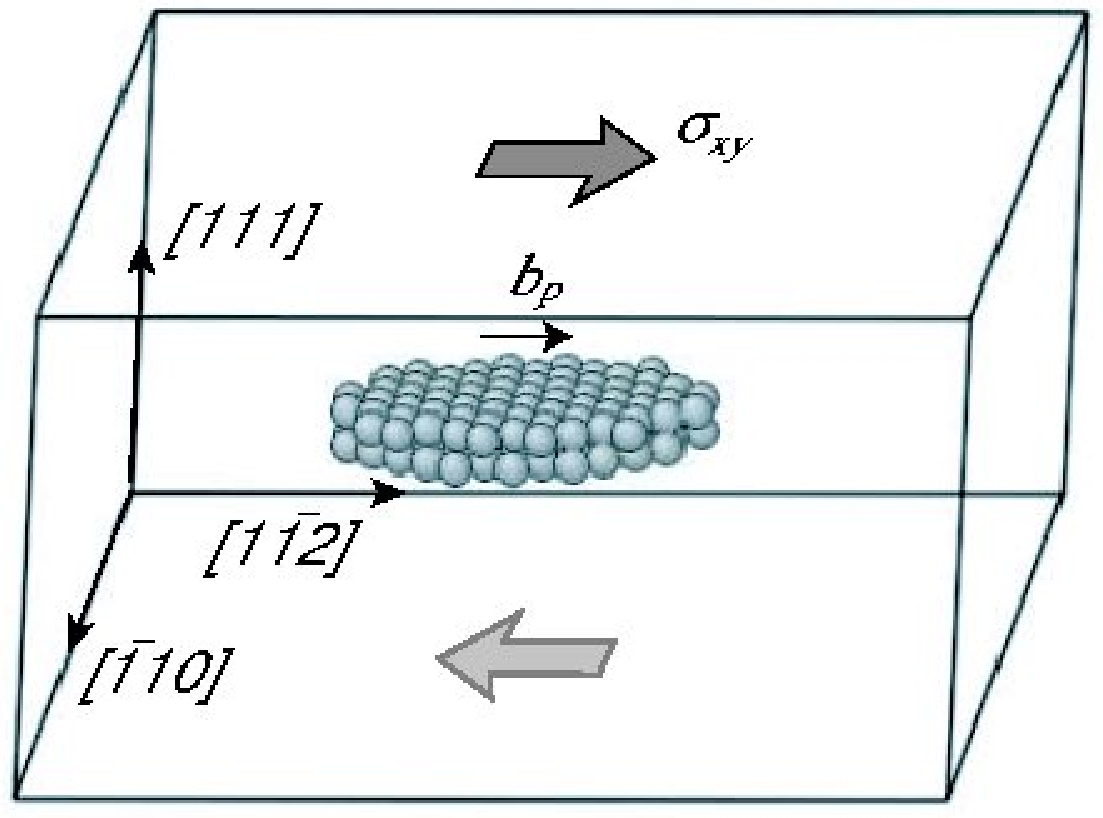}
\includegraphics[width=1.6in]{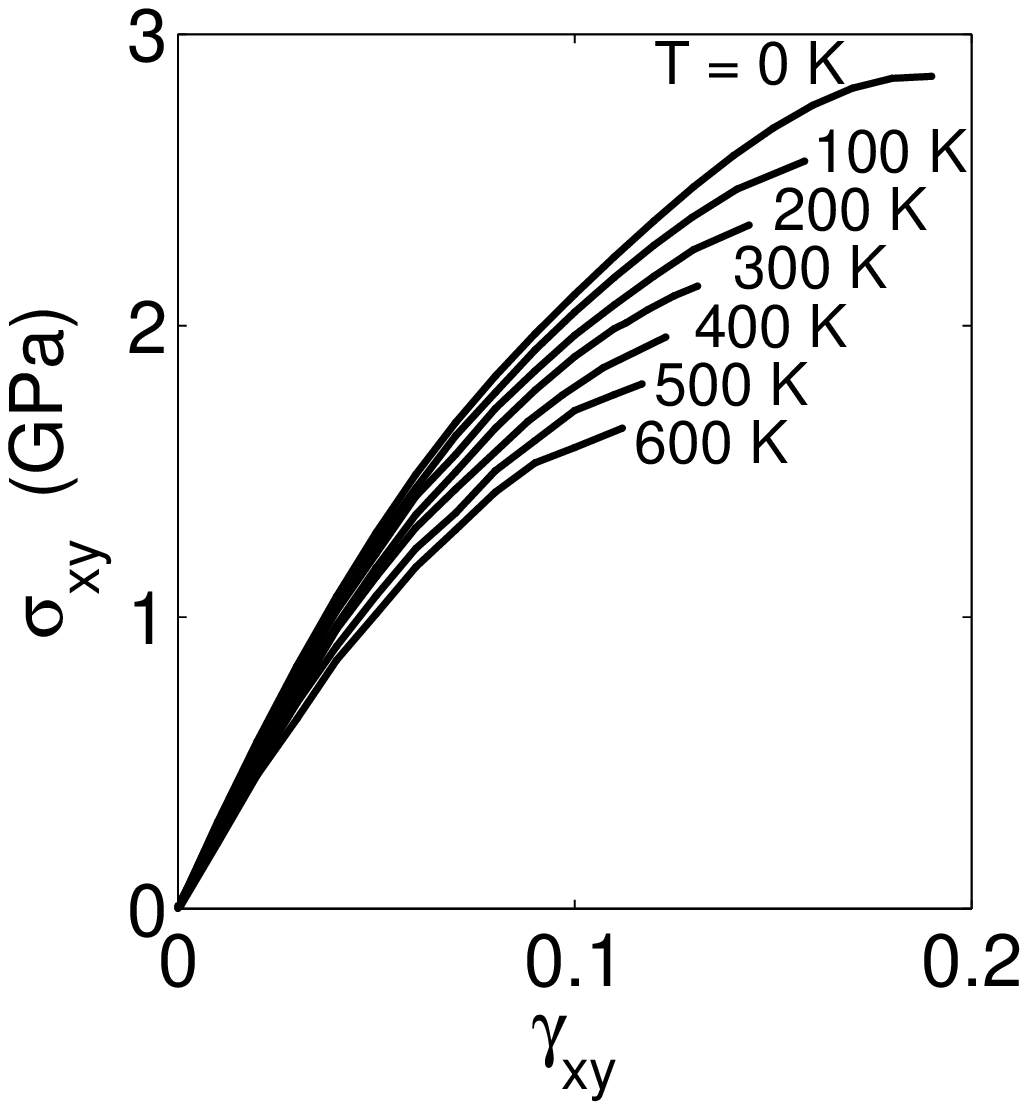} \\
(a)\hspace{1.5in}(b)
\end{center}
\caption{(a) Schematics of the simulation cell. The spheres
represent atoms enclosed by the critical nucleus of a Shockley
partial dislocation loop. (b) Shear stress-strain curves of the Cu
perfect crystal (before dislocation nucleation) at different
temperatures.} \label{fig:crystal}
\end{figure}

For simplicity, we begin with the case of homogeneous dislocation
nucleation in the bulk.  Even though dislocations often nucleate
heterogeneously at surfaces or internal interfaces, homogeneous
nucleation is believed to occur in nano-indentation~\cite{ref:JLi}
and in a model of brittle-ductile transition~\cite{ref:Khantha}.
It also provides an upper bound to the ideal strength of the
crystal.
Our model system is a copper single crystal described by the
embedded-atom method (EAM) potential~\cite{Mishin}.
As shown in Fig.~\ref{fig:crystal}(a), the simulation cell is
subjected to a pure shear stress along $[11\overline{2}]$.
The dislocation to be nucleated lies on the $(111)$ plane and has
the Burgers vector of a Shockley partial~\cite{ref:HL},
$\mathbf{b}_p = [11\overline{2}]/6$.
Fig.~\ref{fig:crystal}(b) shows the shear stress-strain
relationship of the perfect crystal at different temperatures
(before dislocation nucleation).
%
%
%

In this work, we predict the nucleation rate based on the
Becker-D\"oring (BD) theory, which expresses the nucleation rate
per nucleation site as,
\begin{equation}
 I^{\rm BD} = f_c^+\,\Gamma\,\exp\left[
             -\frac{F_c(T,\gamma)}{k_BT}\right]
 \label{eq:IBD}
\end{equation}
where $f_c^+$ is the molecular attachment rate, and $\Gamma$ is
the Zeldovich factor (see Methods).
The BD theory and TST only differs in the frequency prefactor.
Whereas TST neglects multiple recrossing over the saddle point by a
single transition trajectory~\cite{ref:Hanggi}, the recrossing is accounted
for in the BD theory through the Zeldovich factor.


First, we establish the validity of the BD theory for dislocation
nucleation by comparing it against direct MD simulations at a
relatively high stress $\sigma = 2.16$~GPa ($\gamma = 0.135$) at
$T = 300$K, which predicts $I^{\rm MD} = 2.5\times 10^8\,{\rm
s}^{-1}$ (see Methods).
%
%
%
%
The key input to the BD theory is the activation Helmholtz free
energy $F_c(T,\gamma)$, which is computed by umbrella sampling.
%
%
The umbrella sampling is performed in Monte Carlo simulations
using a bias potential as a function of the order parameter $n$,
which is chosen as the number of atoms inside the dislocation loop
(see Methods).

\begin{figure}[tb]
\begin{center}
\includegraphics[width=1.65in]{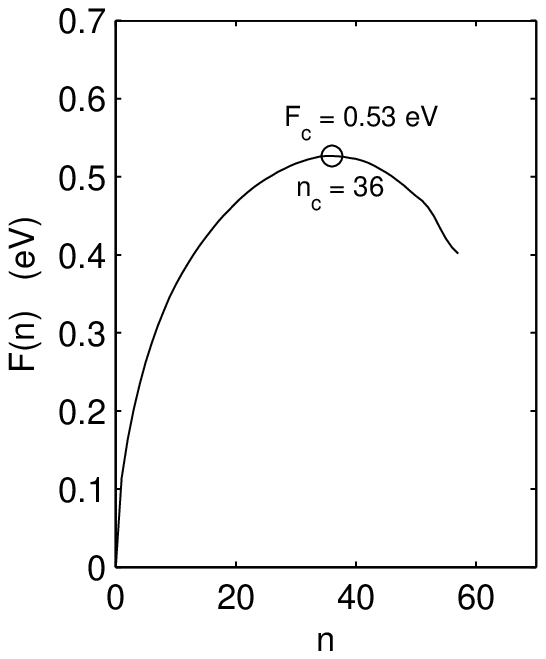}
\hspace{0.05in}
\includegraphics[width=1.65in]{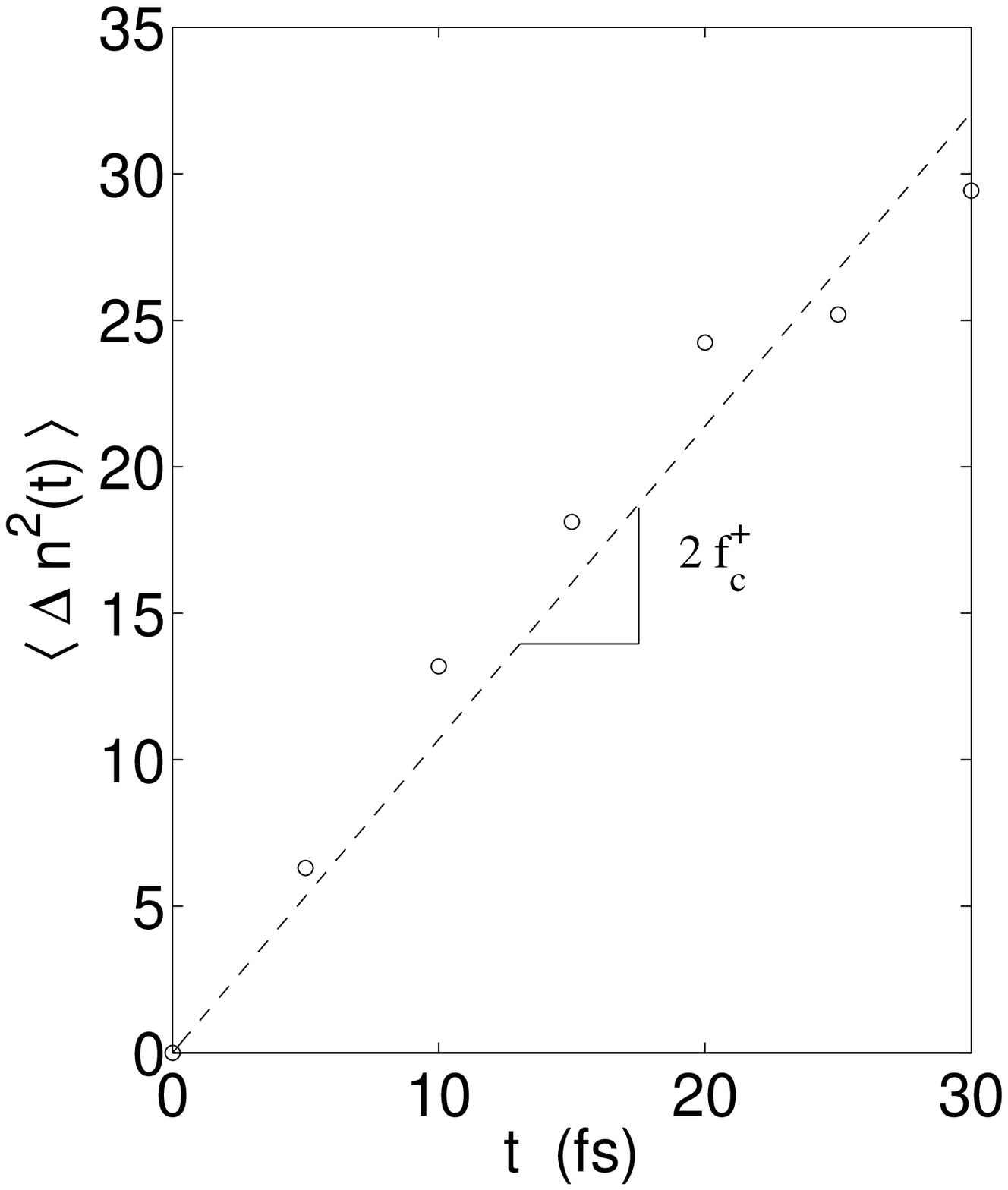} \\
(a)\hspace{1.4in}(b)
\end{center}
\caption{(a) Free energy of dislocation loop during homogeneous
nucleation at $T = 300$~K, $\sigma_{xy} = 2.16$~GPa ($\gamma_{xy}
= 0.135$) from umbrella sampling. (b) Size fluctuation of critical
nuclei from MD simulations. } \label{fig:Gn}
\end{figure}

Fig.~\ref{fig:Gn}(a) shows the free energy function $F(n)$
obtained from umbrella sampling for the specified $(T,\gamma)$
condition.
The maximum of $F(n)$ gives the activation free energy $F_c =
0.53\pm0.01$~eV and the critical nucleus size $n_c = 36$.
The Zeldovich factor~\cite{ref:Zeld}, $\Gamma = 0.055$, is
obtained from $\Gamma \equiv \left( \frac{\eta}{2\pi k_BT}
\right)^{1/2}$ where $\eta = -\left.{\partial^2
F(n)}/{\partial\,n^2} \right|_{n=n_c}$.


Using the configurations collected from umbrella sampling with $n
= n_c$ as initial conditions, MD simulations give the attachment
rate $f_c^+ = 5.3\times10^{14}\,{\rm s}^{-1}$ (see Methods).
Because the entire crystal is subjected to uniform stress, the
number of nucleation sites is the total number of atoms, $N_{\rm
atom} = 14,976$.
Combining these data, the Becker-D\"oring theory predicts the
total nucleation rate to be $N_{\rm atom}I^{\rm BD} = 6.2\times
10^8\,{\rm s}^{-1}$, which is within a factor of 3 of the MD prediction. The
difference between the two is comparable to our error bar.
\myhl{This agreement is noteworthy because no adjustable
parameters (such as the frequency prefactor) is involved in this
comparison. It} shows that the Becker-D\"oring theory and our
numerical approach are suitable for the calculation of dislocation
nucleation rate.

\begin{figure}[t!]
\begin{center}
\includegraphics[width=2.5in]{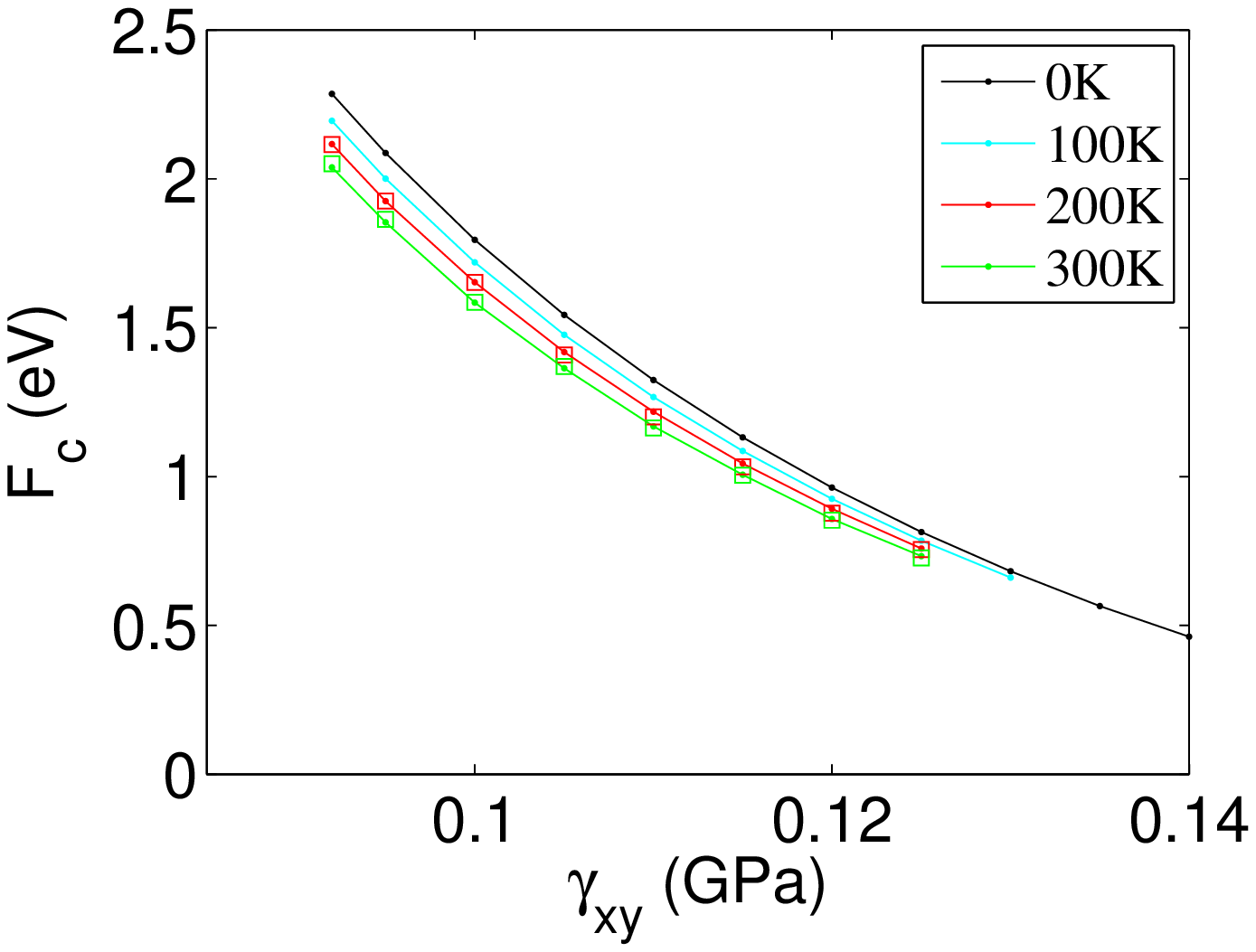} 
\includegraphics[width=2.5in]{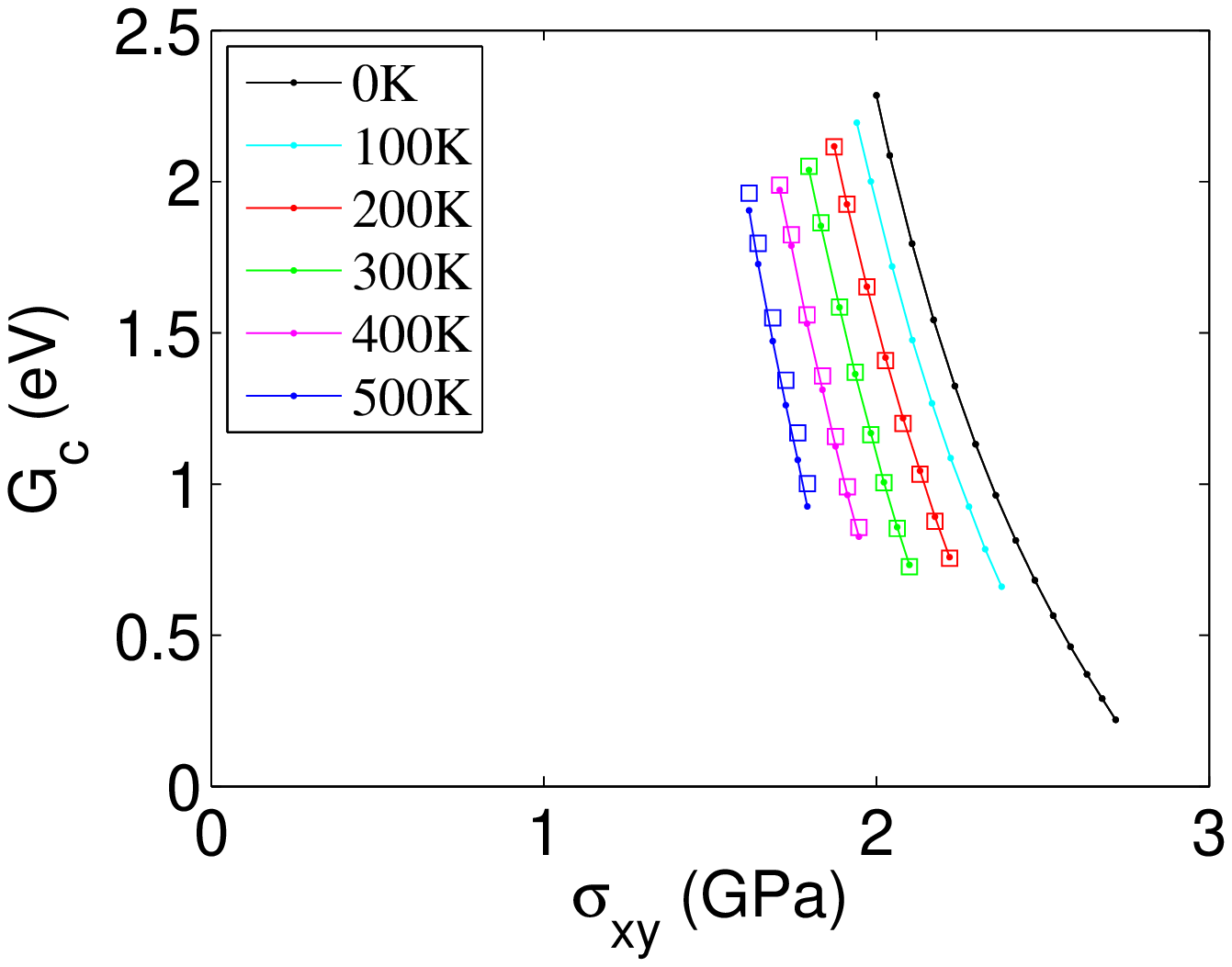} \\
(a)\hspace{2.5in}(b) \\
\includegraphics[width=2.3in]{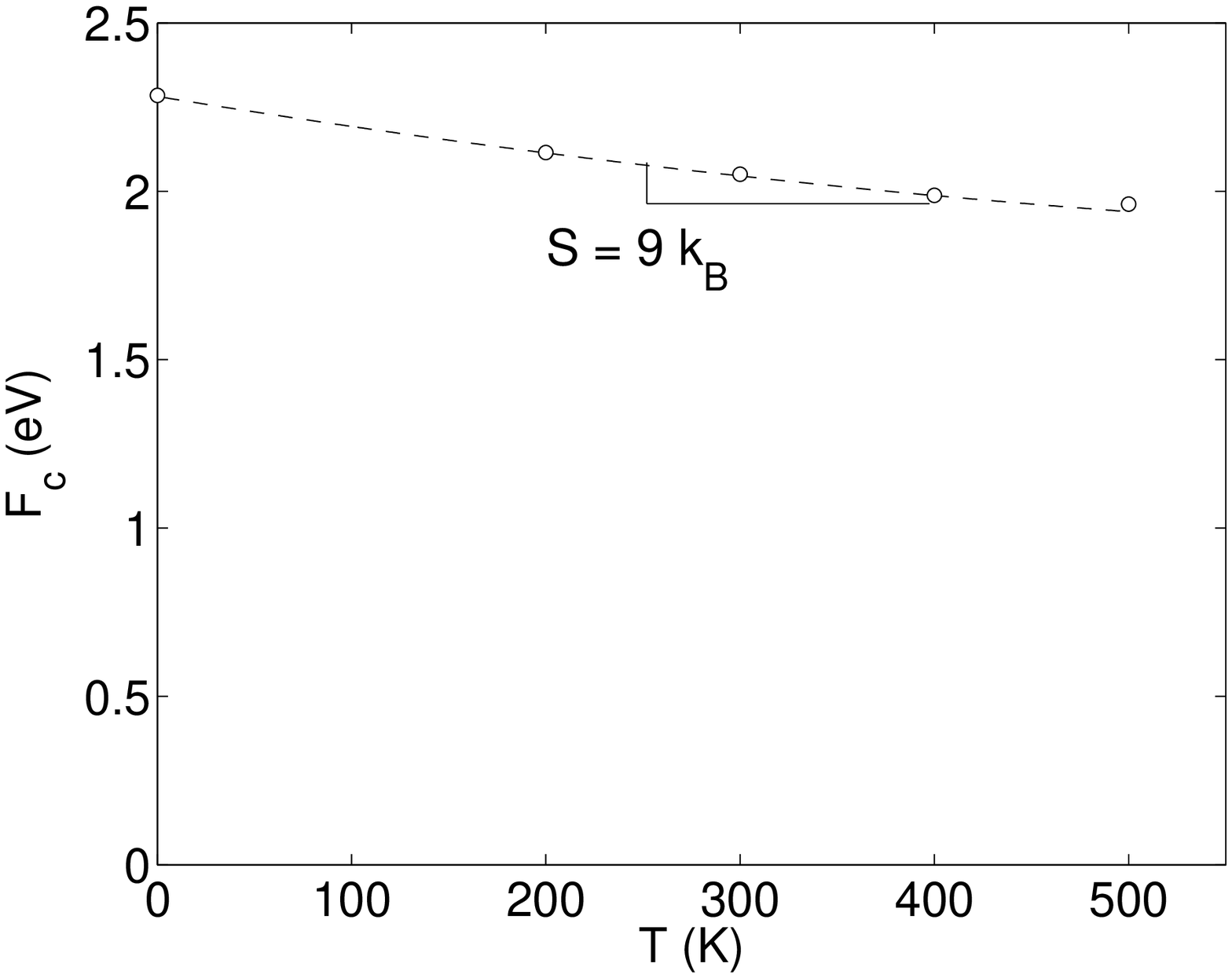}
\includegraphics[width=2.3in]{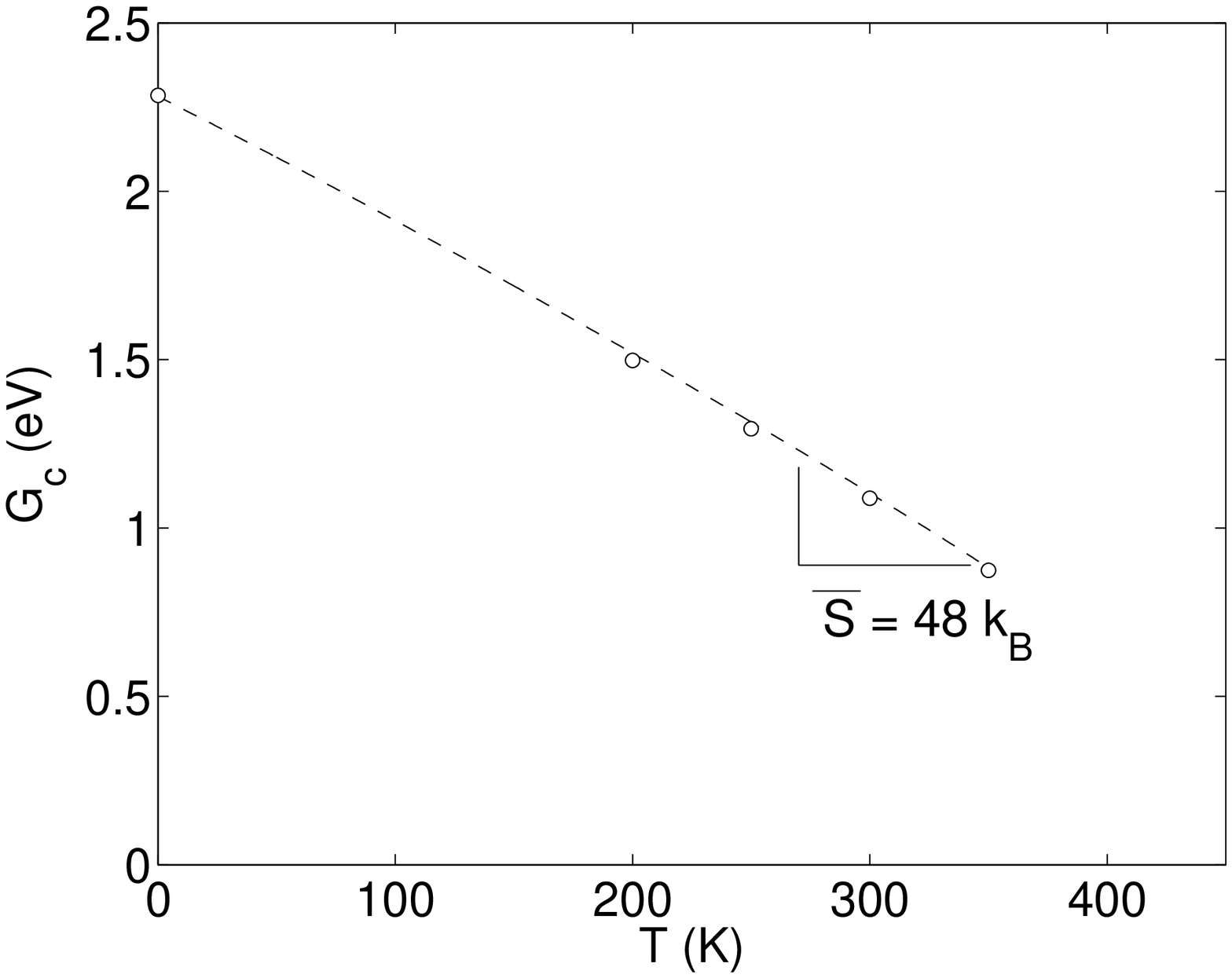} \\
(c)\hspace{2.5in}(d)
\end{center}
\caption{Activation free energy for homogeneous dislocation
nucleation in copper.
(a) $F_c$ as a function of shear strain $\gamma$ at different $T$.
The data for $T=400$K and $T=500$K are shown in SI appendix.
(b) $G_c$ as a function of shear stress $\sigma$ at different $T$.
Squares represent umbrella sampling data and lines represent zero
temperature MEP search results using simulation cells equilibrated
at different temperatures.
(c) $F_c$ as a function of $T$ at $\gamma = 0.092$. Circles
represent umbrella sampling data and dashed line represent a
polynomial fit.
(d) $G_c$ as a function of $T$ at $\sigma = 2.0$~GPa from
polynomial fit.} \label{fig:Gc}
\end{figure}

We now examine the dislocation nucleation rate under a wide range
of temperature and strain (stress) conditions relevant for
experiments and beyond the limited time scale of MD simulations.
Fig.~\ref{fig:Gc}(a) shows the activation Helmholtz free energy
$F_c(T,\gamma)$ as a function of $\gamma$ at different $T$.
The zero temperature data is obtained a minimum-energy-path (MEP)
search using a modified version of the string method, similar to
that used in~\cite{ref:TZhuPNAS,ref:ZhuPRL}.
The downward shift of $F_c$ curves with increasing $T$ is the
signature of the activation entropy $S_c$.
Fig.~\ref{fig:Gc}(c) plots $F_c$ as a function of $T$ at $\gamma =
0.092$.  For $T\leq400$K, the data closely follow a straight line, whose slope
gives $S_c(\gamma) = 9~k_B$.
This activation entropy contributes a significant multiplicative factor,
$\exp(S_c/k_B) \approx 10^4$, to the absolute nucleation rate, and
cannot be ignored.
What causes this rapid drop of activation free energy with
temperature?  Thermal expansion and vibrational entropy are two
candidate mechanisms.
To examine the effect of thermal expansion, we performed zero
temperature MEP search at $\gamma = 0.092$, but with other strain
components fixed at the equilibrated values at $T = 300$~K.
\myhl{This approach is similar to the quasi-harmonic approximation
(QHA)}~\cite{ref:Foiles1994,ref:deKoning2002} \myhl{often used in
free energy calculations in solids, except that, unlike QHA, the
vibrational entropy is completely excluded here.}
The resulting activation energy, $\tilde{E_c} = 2.04$~eV, is
indistinguishable from the activation free energy $F_c=2.05 \pm 0.01$ at $T =
300$~K computed from umbrella sampling.
Because atoms do not vibrate in the MEP search, this result shows
that the dominant mechanism for the large $S_c(\gamma)$ is thermal
expansion, while the contribution from vibrational entropy is
negligible.
As temperature increases, thermal expansion pushes neighboring
atoms further apart and weakens their mutual interaction.
This expansion makes crystallographic planes easier to shear and
significantly reduces the free energy barrier for dislocation
nucleation.
In the widely used harmonic approximation of TST, the activation
entropy is often attributed to the vibrational degrees of freedom
as $\nu_0 \exp(S_c/k_B) = (\prod_{i=1}^{\cal N} \nu_i^{\rm m}) /
(\prod_{i=1}^{{\cal N}-1} \nu_i^{\rm a})$, where $\nu_i^{\rm m}$
and $\nu_i^{\rm a}$ are the positive normal frequencies around the
local energy minimum and activated state,
respectively~\cite{ref:Hanggi,ref:Vineyard,ref:Gupta2002}.
%
%
However, here we see that $S_c(\gamma)$ arises entirely from the
anharmonic effect for dislocation nucleation.
At $T=400$K and $T=500$K, we observe significant differences between
$F_c$ computed from umbrella sampling and $\tilde{E_c}$ computed from
MEP search in the expanded cell. 
These difference must also be attributed to anharmonic effects.
While it is easier to control strain $\gamma$ than stress $\sigma$
in atomistic simulations, it is usually easier to apply stress in
experiments, and experimental results are often expressed as a
function of $\sigma$ and $T$.
To bridge between simulations and experiments, it is important to
establish a connection between the constant-stress and
constant-strain ensembles.
In the constant-strain ensemble, the system is described by the
Helmholtz free energy $F(n,T,\gamma)$ where $n$ is the size of the
dislocation loop and the activation Helmholtz free energy is
defined as $F_c(T,\gamma) \equiv F(n_c,T,\gamma) -
F(n\!\!=\!\!0,T,\gamma)$.
In the constant-stress ensemble, the system is described by the
Gibbs free energy $G(n,T,\sigma)$, from the Legendre transform
$G=F-\sigma\,\gamma\,V$, with $\sigma \equiv V^{-1}\partial
F/\partial \gamma|_{n,T}$.  Similarly, $G_c(T,\sigma) \equiv
G(n_c,T,\sigma) - G(n\!\!=\!\!0,T,\sigma)$.
We have proved that $G_c(T,\sigma) = F_c(T,\gamma)$ in the
thermodynamic limit of $V\to\infty$, when $\sigma$ and $\gamma$
satisfies the stress-strain relation of the perfect crystal,
$\sigma(\gamma,T)$. The difference between $F_c$ and
$G_c$ when $\sigma=\sigma(T,\gamma)$ is of the order ${\cal
O}(V^{-1})$. The details of the proof will be published
separately.

Combining the activation Helmholtz free energy $F_c(T,\gamma)$
shown in Fig.~\ref{fig:Gc}(a) and the stress-strain relations
shown in Fig.~\ref{fig:crystal}(b), we obtain the activation Gibbs
free energy $G_c(T,\sigma)$, which is shown in
Fig.~\ref{fig:Gc}(b).
We immediately notice that the curves at different temperatures
are more widely apart in $G_c(T,\sigma)$ than that in
$F_c(T,\gamma)$, indicating a much larger activation entropy in
the constant-stress ensemble.
%
%
%
For example, Fig.~\ref{fig:Gc}(d) plots $G_c$ as a function of $T$
at $\sigma = 2.0$~GPa, from which we can obtain an averaged
activation entropy of $\overline{S_c}(\sigma)=48\,k_B$ in the
temperature range of $[0, 300{\rm K}]$.
%
%
This activation entropy contributes a multiplicative factor of
$\exp(\overline{S_c}(\sigma)/k_B) \approx 10^{20}$ to the absolute
nucleation rate.

The dramatic difference between $S_c(\gamma)$ and $S_c(\sigma)$
may seem surprising.
Indeed, they are sometimes used
interchangeably~\cite{ref:Huntington55,ref:DiMelfi76}, although
the conceptual difference between the two has been pointed out in
the context of chemical
reactions~\cite{ref:Whalley64,ref:Tonnet75}.
It is well known that the entropy is independent of the ensemble
of choice, i.e. $S(n,T,\gamma) \equiv \partial
F(n,T,\gamma)/\partial T|_{n,\gamma}$ and $S(n,T,\sigma) \equiv
\partial G(n,T,\gamma)/\partial T|_{n,\sigma}$ equal to each other
as long as $\sigma = V^{-1}\partial F/\partial \gamma|_{n,T}$,
which is true by definition.
At the same time, the activation entropy is just the entropy
difference between the activated state and the metastable state,
i.e. $S_c(T,\gamma) = S(n_c,T,\gamma)-S(n\!\!=\!\!0,T,\gamma)$ and
$S_c(T,\sigma) = S(n_c,T,\sigma)-S(n\!\!=\!\!0,T,\sigma)$.
If the entropies in two ensembles can equal each other, it may
seem puzzling how the activation entropies can be different.

The resolution of this apparent paradox is that under the constant
applied stress, the nucleation of a dislocation loop causes a
strain increase, i.e. $\sigma(n\!\!=\!\!0,T,\gamma) =
\sigma(n_c,T,\gamma^+)$, with $\gamma^+ > \gamma$.
Based on this result, one can show that the difference in the
activation entropies, $\Delta S_c \equiv S_c(\sigma) -
S_c(\gamma)$, equals
$S(n\!\!=\!\!0,T,\gamma^+)-S(n\!\!=\!\!0,T,\gamma)$, which is the
entropy difference of the perfect crystal at two slightly
different strains.
We can further show that $\Delta S_c = -\Omega_c(\sigma)\partial
\sigma/\partial T|\gamma$, where $\Omega_c\equiv -\partial
G_c/\partial \sigma|_T$ is the activation volume and $-\partial
\sigma/\partial T|\gamma$ describes the thermal softening effect.
Hence $\Delta S_c$ is always positive for nucleation processes in
solids driven by shear stress.
In the case of homogeneous dislocation nucleation $\Delta S_c$ as
large as $39\,k_B$ is observed for homogeneous dislocation
nucleation, which is mainly caused by its large activation volume
$\Omega_c$.
The numerical results enable us to examine the
approximation~\cite{ref:ZhuPRL} based on the so-called
thermodynamic ``compensation law''~\cite{ref:Kemeny73},
%
%
which states that the activation entropy is proportional to the
activation enthalpy (or energy).
We find that $S_c(\gamma)$ can be roughly approximated by
$E_c(\gamma)/T^*$ with $T^*\approx3000$~K while $S_c(\sigma)$ is
not proportional to $H_c(\sigma)$ (see SI appendix).
%
%

%
%
%


\begin{figure}[ht!]
\begin{center}
\includegraphics[height=1.4in]{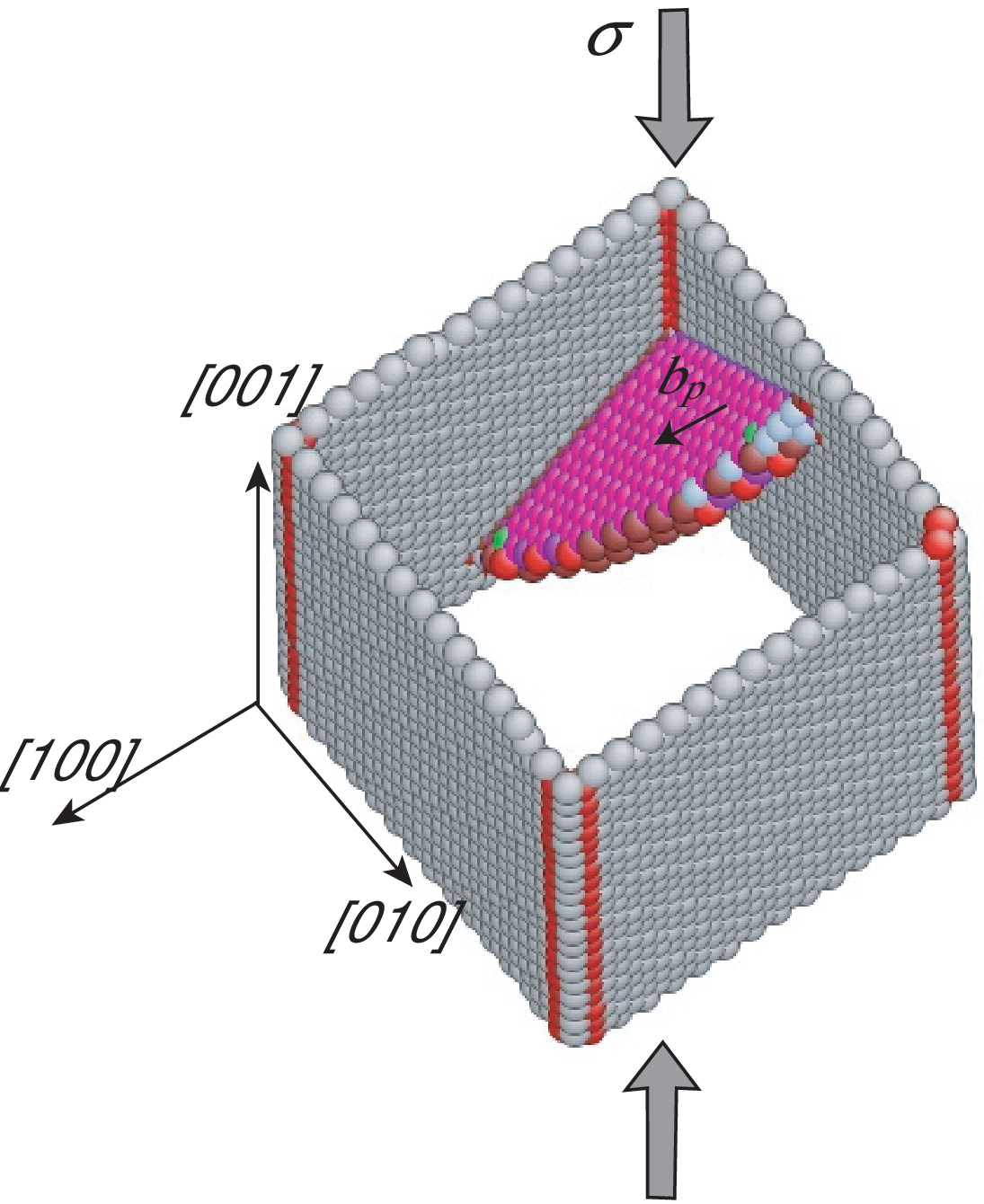}
\includegraphics[height=1.5in]{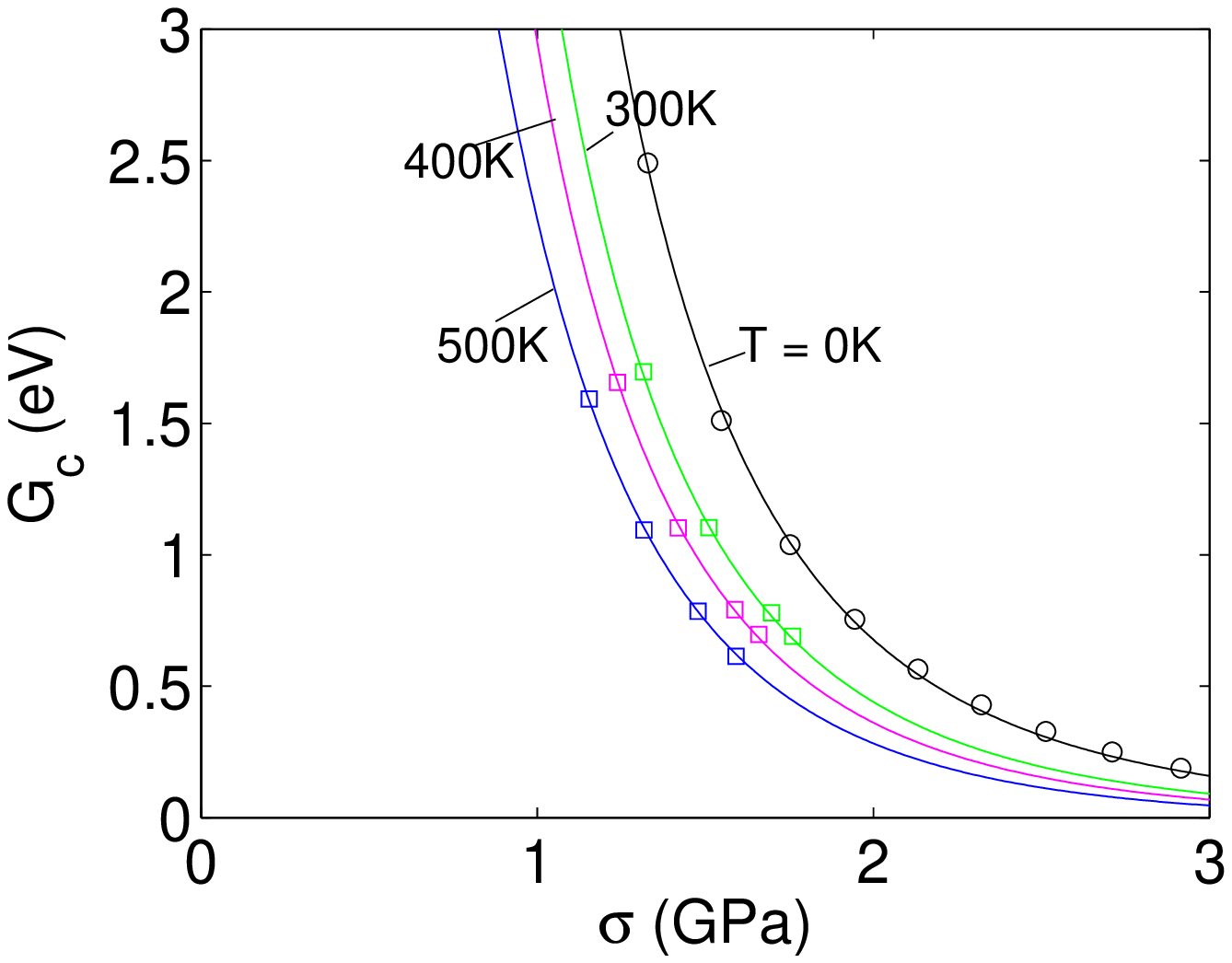} \\
(a)\hspace{1.5in}(b)\\
\includegraphics[height=2.0in]{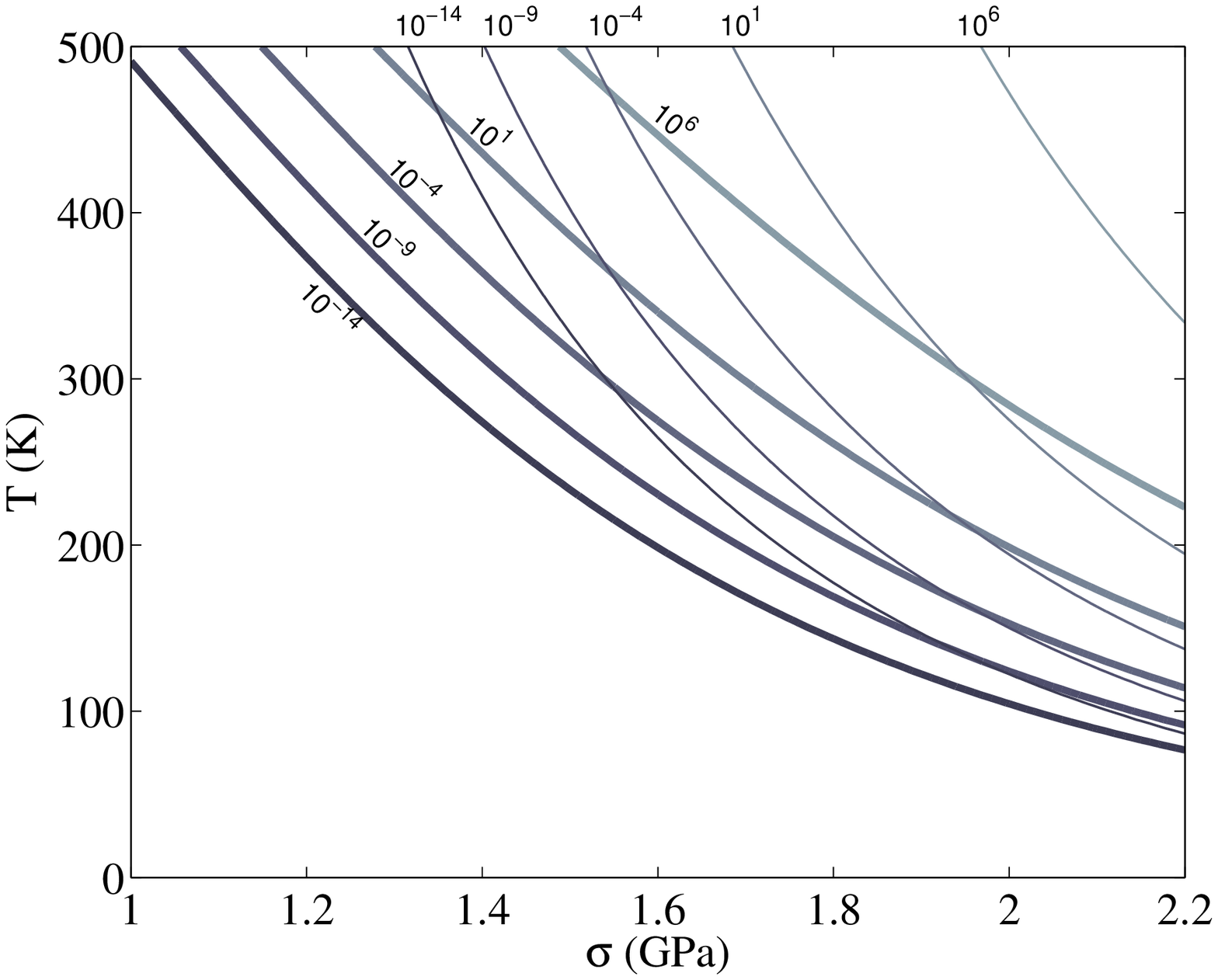} \\
(c)
\end{center}
\caption{(a) Heterogeneous dislocation nucleation in a copper
nanorod under compression, visualized by Atomeye~\cite{ref:Atomeye}.
(b) $G_c$ as a function compressive stress $\sigma$ at different
$T$.
(c) Contour lines of dislocation nucleation rate per site $I$ as a
function of $T$ and $\sigma$.  The predictions with and without
accounting for activation entropy $S_c(\sigma)$ are plotted in
thick and thin lines, respectively.
The nucleation rate of $I\sim10^{6}{\rm s}^{-1}$ per site is
accessible in typical MD time scales while nucleation rate of
$I\sim10^{-4}$-$10^{-9}$ is accessible in typical experimental
time scales, depending on the number of nucleation sites. }
\label{fig:Gc_hetero}
\end{figure}

To assess the applicability of these conclusions in heterogeneous
nucleation, we studied dislocation nucleation from the corner of a
$[001]$-oriented copper nanorod with $\{100\}$ side surfaces under
axial compression (see Methods).
Fig.~\ref{fig:Gc_hetero}(b) plots the activation free energy
barrier as a function of axial compressive stress $\sigma$, which
shows significant reduction of the activation free energy with
temperature.
For example, at the compressive elastic strain of $\epsilon=0.03$,
the compressive stress is $\sigma = 1.50$~GPa at $T = 0$~K.
The activation entropy $S_c(\epsilon)$
at this elastic strain equals $9 k_B$, whereas the activation
entropy $S_c(\sigma)$
at this stress equals $17 k_B$.
Fig.~\ref{fig:Gc_hetero}(c) plots the contour lines of the
predicted dislocation nucleation rate (per nucleation site) as a
function of $T$ and $\sigma$.
To show the physical effect of the large activation entropies, the
dashed lines plot the rate predictions if the effect of
$S_c(\sigma)$ were completely neglected.
Significant deviations between the two sets of contour lines are
observed, especially for $T\geq 300$~K and $\sigma\leq 1.5$~GPa.
For example, at $T=300$~K and $\sigma=1.5$~GPa (where a thick and
a thin contour line cross), the neglect of activation entropy
would cause an underestimate of the nucleation rate by 10 orders
of magnitude.

\comment{
To examine whether large activation entropies exist in other
nucleation processes in solids driven by shear stress, we study
the cross slip of a screw dislocation in nickel modelled by the
EAM potential~\cite{ref:EAMNi}.
Cross slip is well recognized as the fundamental process
controlling a wide range of mechanical properties, including
strain hardening, dynamic recovery and fatigue~\cite{ref:Puschl}.
To be specific, we consider the effect of the shear stress
$\sigma$ that only couples to the edge component of the partial
dislocations on the cross slip plane and exert zero stress on the
partial dislocations on the original glide plane, and let $\gamma$
be the corresponding shear strain (see Methods).
%
%
%
%
By evaluating the activation volume $\Omega_c$ and thermal
softening effect $-\partial \sigma/\partial T|_\gamma$, we find
that at $\sigma = 500$~MPa, $\Delta S_c \equiv S_c(\sigma) -
S_c(\gamma) \approx 7 k_B$.
%
%
%
}

In summary, we have shown that the Becker-D\"oring theory combined
with free energy barriers determined by umbrella sampling can
accurately predict the rate of homogeneous dislocation nucleation.
In both homogeneous and heterogeneous dislocation nucleation, a
large activation entropy at constant elastic strain is observed,
and is attributed to the weakening of atomic bonds due to thermal
expansion.
An even larger activation entropy is observed at constant stress,
due to thermal softening.
Both effects are anharmonic in nature, and emphasize the need to
go beyond harmonic approximation in the application of rate
theories in solids.
%
%
%
We believe our methods and the general conclusions are applicable
to a wide range of nucleation processes in solids that are driven
by shear stress, including cross slip, twinning and martensitic
phase transformation.

\begin{description}
\item[Methods]

\item[Molecular Dynamics]
%
The simulation cell for homogeneous dislocation nucleation has
dimension $8[11\overline{2}]\times 6[111]\times
3[1\overline{1}0]$.
%
%
Periodic boundary conditions (PBC) are applied to all 3
directions.
To reduce artifacts from periodic image interactions, the
applied stress is always large enough so that the diameter of
critical dislocation loop is smaller than half the width of the
simulation cell.

The shear strain $\gamma$ is the $x$-$y$ component of the
engineering strain.
%
%
The following procedure is used to obtain pure shear stress-strain
curve shown in Fig.~\ref{fig:crystal}(b).
At each temperature $T$ and shear strain $\gamma_{xy}$, a series
of $2$ ps MD simulations under the canonical, constant temperature-constant volue (NVT) ensemble are performed.
After each simulation, all strain components except $\gamma_{xy}$
are adjusted according to the average Virial stress until
$\sigma_{xy}$ is the only non-zero stress component.
The shear strain is then increased by $0.01$ and the process
repeats until the crystal collapses spontaneously.
The shear stress-strain data are fitted to a polynomial function,
$\sigma(\gamma,T)=\sum_{i=0}^{2}\sum_{j=0}^{2}a_{ij}\gamma^iT^j$.
To obtain average nucleation time at $\sigma_{xy}=2.16$ GPa
($\gamma=0.135$) at $300$~K, we performed $192$ independent MD
simulations using the NVT ensemble with random initial velocities.
Each simulation runs for $4$~ns.  If dislocation nucleation occurs
during this period, the nucleation time is recorded.  This
information is used to construct the function $P_s(t)$, which is
the fraction of MD simulation cells in which dislocation
nucleation has not occurred at time $t$.
$P_s(t)$ can be well fitted to the form of $\exp(-I^{\rm MD}t)$ to
extract the nucleation rate $I^{\rm MD}$.

To compute the attachment rate $f_c^+$, we collect from umbrella
sampling an ensemble of $500$ atomic configurations for which
$n=n_c$, and run MD simulations using each configuration as an
initial condition.  The initial velocities are randomized
according to Boltzmann's distribution.
The mean square change of the loop size, $\langle \Delta
n^2(t)\rangle$, as shown in Fig.~\ref{fig:Gn}(b), is fitted to a
straight line, $2 f_c^+ t$, in order to extract
$f_c^+$~\cite{ref:RyuCai}.

\item[ Free energy barrier calculations ]
%
The reaction coordinate $n$ is defined for each atomic
configurations in the following way.
An atom is labelled as ``slipped'' if its distance from any of its
original nearest neighbors has changed by more than the critical
distance $d_c$~\cite{ref:Ngan}.
We choose $d_c = 0.33\text{\AA}$, $0.38\text{\AA}$ and $0.43\text{\AA}$ for $T\leq 400$~K,
$T=500$~K and $T = 600$~K, respectively.
%
%
The ``slipped'' atoms are grouped into clusters; two atoms belong
to the same cluster if their distance is less than cutoff distance
$r_c$~($3.4\text{\AA}$).
The reaction coordinate $n$ is the number of atoms in the largest
cluster divided by two.
%
%
%
%

To perform umbrella sampling, a bias potential
$k_B\,\hat{T}\,(n-\overline{n})^2$ is superimposed on the EAM
potential, where $\hat{T}=40\,{\rm K}$ and $\overline{n}$ is the
center of the sampling window.
We choose $\hat{T}$ empirically  so that the width of the sampling
window on the $n$-axis is about 10.
%
%
%
The activation Helmholtz free energy for homogeneous nucleation data can be fitted very well by a
polynomial function,
$F_c(\gamma,T)=\sum_{i=0}^{2}\sum_{j=0}^{2}b_{ij}\gamma^iT^j$
in the range of $0.09 \leq \gamma \leq 0.12$ and $0 \leq T \leq 500$K.
%

%

For heterogeneous dislocation nucleation, the size of the copper
nanorod~\cite{ref:ZhuPRL} is $15[100]\times 15[010]\times 20[001]$
with PBC along $[001]$. The activation Gibbs free energy for heterogeneous
nucleation is fitted to an empirical form $G_c(\sigma,T)=A( (\sigma/\sigma_0)^p-1)^q-B\sigma^l T$.
The error bar of activation free energies is about $0.5 k_B T$. Hence, the error bar
of activation entropies is about $0.5 k_B$.

\end{description}

\acknowledgments
This work is supported by the National Science Foundation Grant CMS-0547681, 
a Department of Energy Scienctific Discovery through Advanced Computing project 
on Quantum Simulation of Materials and Nanostructures, and the Army High Performance Computing Research Center
at Stanford. S.R. acknowledge the support from the Weiland Family
Stanford Graduate Fellowship.

\end{document}